# Investigation of Magnetic Domain Structure in $Bi_{0.7}Dy_{0.3}FeO_3$ thin films integrated with ZnO film

Deepak Bhatia[1,a)], Himanshu Sharma[2,b)], Seeraj Nawaj[1], R.S.Meena[3], C.V. Tomy[2], V.R Palkar[1]

[1] *Department of Electrical Engineering and Centre for Excellence in Nanoelectronics, Indian Institute of Technology Bombay, Mumbai-400076, India.*
[2] *Department of Physics, Indian Institute of Technology Bombay, Mumbai-400076, India.*
[3]*Department of Electronics Engineering, Rajasthan Technical University, Kota-324010, India.*

**Abstract:** The magnetic domain structure of the multiferroic $Bi_{1-x}Dy_xFeO_3$ (BDFO) deposited on ZnO at the macroscopic level has been demonstrated in this paper. Magnetic properties are observed by saturated magnetic and ferroelectric hysteresis loops at room temperature. The coupling behaviour and magnetic transition are verified using Multimode Atomic Force Microscope by applying bias between sample and MFM tip. ZnO thin film of 300 nm was deposited by dielectric sputtering using a ZnO target on Si p-type conducting (.0001-0005 $\Omega$ cm) substrate. BDFO thin films of 300 nm were deposited on Si/ZnO using PLD (Pulsed laser deposition) technique. It is observed that BDFO thin films integrated with ZnO film shows the coexistence of ferromagnetic and ferroelectric ordering with significant coupling at room temperature. Integration of BDFO films with ZnO piezoelectric thin film shows the potential in MEMS applications as well as in Memory devices and strong history dependent systems.

## 1. INTRODUCTION

Multiferroics having coexistence of ferroelectric and ferromagnetic ordering in the same phase at room temperature have large potential for variety of device applications. Recently developed magnetoelectric multiferroic "Dy" modified $BiFeO_3$ ($Bi_{1-x}Dy_xFeO_3$ or BDFO) is new material exhibiting the ideal properties at room temperature. Furthermore, BDFO shows the significant coupling, i.e., ferroelectric and ferromagnetic ordering at room temperature simultaneously [1-2]. However, the presence of magnetoelectric (M–E) coupling between electric and magnetic order parameters has been theoretically predicted [6]. Hence, there is intense interest in its implementation in device architectures by taking advantage of the said properties [11]. Among the known one dimensional nano materials, Zinc oxide (ZnO) is unique because of its semiconductor and piezoelectric dual properties. It finds applications in MEMS due to unique combinations of electrical, optical and piezoelectric properties. However ZnO is sensitive material for wet etching and treatment by temperature, acid bases and even water [3-4]. The film of BDFO using the Pulse Laser Deposition(PLD) on p-type Si film are also reported, it exhibits excellent dielectric properties such as high dielectric constant (k = 30), low leakage and low interface state density may provide the storage of generated electricity as memory device [6]. The pervoskite $BiFeO_3$ exhibits antiferrromagnetic ($T_N \sim 380^oC$) and ferroelectric (Tc $\sim 810^\circ C$) ordering, which ensures BFO to be the most promising candidate for high temperature device applications [6-9]. It is essential to study the Magnetic domain structure, in the interest of using such remarkable properties. In this paper, we report the M-H coupling at room temperature and saturation observed on magnetic hysteresis loop. Coupling is also verified by observing the effect of voltage on magnetic domain pattern. A Multimode Atomic Force Microscope in MFM mode has been used for the study and images are obtained on specific spatial area. Ferroelectric hysteresis loop (P-E) obtained at room temperature by loop tracer Precision LC (Radiant Technology, USA) in order to determine the effect of magnetic field on ferroelectric properties of BDFO and BDFO/ZnO thin films [13]. However to the best of our knowledge the characteristics of multiferroic thin film integrated with piezoelectric directly deposited on p-type low resistivity Si for storage purpose of electricity generated by BDFO/ZnO nanocomposite have not been reported. The excellent multifunctional properties of BDFO/ZnO films could find for a variety of device applications like sensors, energy scavengers, electricity generators, etc [2].

## 2. EXPERIMENTAL

ZnO thin films were deposited on Si by dielectric sputtering using a ZnO target (99.9%) with a diameter of 2 inch and thickness 3 mm. Conducting p-type Si substrate with resistivity of 0.0001 $\Omega$-cm was used for deposition. In order to avoid the dead layer, standard RCA (Radio

Email: b)himsharma@iitb.ac.in
a) keshav_eck26@yahoo.com

Corporation of America) cleaning was performed on silicon wafers before loading them into dielectric sputter. During the deposition of ZnO thin films the RF power was 150 W, the base pressure was $5\times10^{-5}$ mbar and operating pressure was $2.2\times10^{-2}$ mbar. Thin films were deposited in Ar-atmosphere with a deposition rate of 15 nm/min.

On the other hand powder sample of $Bi_{0.7}Dy_{0.3}FeO_3$ was prepared using partial co-precipitation route described elsewhere [6]. The powder material was compacted in the pellet form and sintered at 800°C for 2 hours [8]. This highly dense pellet thus obtained was used as target for Pulsed laser deposition (PLD) technique. BDFO thin films of 300 nm were deposited using PLD. Complex Pro 201, KrF excimer laser with $\lambda = 248$ nm and energy density [6] of 2 J/cm², was used for deposition of BDFO thin films. During deposition target to substrate distance, substrate temperature and $O_2$ Pressure were 5 cm, 650°C and $4.5\times10^{-4}$ mbar respectively. Thickness of thin film was found to be 300 nm using the 22500 number of pulses with a repetition rate of 10 Hz. The thickness of the film was estimated by profilometer (Ambios, USA). Further, the characterization of these films was done by various techniques. Phase purity and crystal structure were determined by X-ray diffraction. However, the surface morphology of the films was carried out using a Multimode Atomic Force microscope (MAFM) in MFM mode, scanning electron microscopy (SEM) and atomic force microscope (AFM). Magnetization of the thin films, as a function of applied magnetic field at room temperature was measured using a SQUID-VSM (Quantum Design Inc., USA) and electric polarization measurements were carried out using ferroelectric hysteresis loop tracer Precision LC (Radiant Technology, USA).

## 3. RESULTS AND DISCUSSION

### 3.1. Crystal Structure and Surface Morphology

XRD Patterns obtained for ZnO and BDFO/ZnO films are shown in Fig.1. X-ray from Rigaku (Cu-Kα radiation, $\lambda = 1.5405$Å) was used for structural phase identifications. The heterostructure (BDFO/ZnO) exhibit a pervoskite structure similar to that of pure BFO. It is obvious that Dy substitution has affected the structure of parent compound BFO. All the peaks of the patterns are indexed.

The XRD pattern indicates that the BDFO film deposited on the ZnO films is a single phase and polycrystalline in nature. This is due to the lattice mismatch between ZnO and BDFO. However, the XRD pattern of ZnO thin film indicates that diffraction peak located around $34.422^0$ is very high and just ZnO (002) diffraction peak is observed. So the deposited ZnO thin film on the Si substrate has a high c-axis preferred orientation, which is essential to achieve a ZnO film with high piezoelectric quality [3-4].

Scanning electron microscopy (SEM) was done using Raith-150 to determine the grain morphology of BDFO and ZnO films. SEM is also used to find the uniformity and the thickness of BDFO films over ZnO films. SEM images reveal the BDFO coating with granular structure (See Fig. 2). Fig. 2 Shows the SEM images of ZnO and BDFO/ZnO films at different magnifications.

Further, Fig. 3 and Fig. 4 shows atomic force microscope (AFM) images of BDFO thin film and BDFO/ZnO thin film. A multimode atomic force microscope (Nanoscope IV from Digital Instruments) configured to provide the magnetic domain structure has been used in this study. The images are obtained on a specific spatial area of the film spanning tens of microns on the lateral scale with a spatial resolution better than 50 nm. Additionally, the magnetic force microscopy (MFM) images were obtained by using a tapping cantilever with a cobalt-coated tip that has been magnetized with a strong permanent magnet before installing on the AFM head. A tapping-mode topographic image was obtained during the main scan in order to investigate any correlation between the local order parameter and the sample morphology [8], [12]. During the interleave mode, the tip is raised above the sample surface, allowing the imaging of relatively weaker but long-range magnetic interactions while minimizing the influence of topography [11]. The resulting images from the raw data for topography and magnetic structure are shown in figures without any additional image processing. The topological (AFM) images of BDFO/ZnO and BDFO films respectively with scan size 500nm x 500nm, 1μm x 1μm and color scale at 50nm are shown in Fig. 3 and Fig. 4. Fig. 5 and Fig. 6 included the MFM images of BDFO and BDFO/ZnO films, respectively with scan size 500nm x 500nm, 1μm × 1μm. Phase image of BDFO/ZnO film is shown in Fig. 7. This reflects the presence of magnetic stripe domains in the BDFO and BDFO/ZnO thin films. The magnetic image shows a highly irregular zigzag domain pattern. The domain size is five to ten times larger than the typical grain size on the linear scale, as seen from topography images. Fig. 3-7 Multimode scanning probe images (topography and magnetic domain) of BDFO/ZnO thin film: The images show unambiguously the ferroelectric writing and the existence of remnant polarization reflects the presence of magnetic stripe domains in the BDFO and BDFO/ZnO thin films.

### 3.2. Magnetic Properties

Fig. 8 shows the variations of magnetization with magnetic field (M-H curve) for the BDFO, ZnO and BDFO/ZnO films at room temperature respectively. The inset of Fig. 8a, 8b and 8c highlights the coactivity of corresponding thin films. It is observed that the Dy enhance the magnetization of $Bi_{1-x}Dy_xFeO_3$ in compare that of BFO [9], [22]. However, the saturation magnetic field is found to be small

*Email:* ᵇ⁾himsharma@iitb.ac.in
ᵃ⁾keshav_eck26@yahoo.com

in compare to BFO. The small field requirement to bring saturation indicates that magnetic domains could easily be polarized at low fields. This property could be advantageous during device operation. Further, an enhanced magnetization and enhanced saturation magnetic field is observed in in BDFO/ZnO system at room temperature in compare of BDFO. There are various possible reasons that are known to cause magnetic anisotropy in the films, for example, magneto crystalline structure of the material and texturing, stress during growth of the film, grain size (crystalanity) etc. Out of magnetocrystalline anisotropy is an intrinsic property of a magnetic material independent of the grain size and shape [7]. Depending on the crystallographic orientation of the sample in the magnetic field, the magnetization reaches saturation value at different fields.

### 3.3. Ferroelectric Properties

The main objective of the measurement was to study the electric field on the nature of P-E curve, saturation and remnant polarization. In order to determine the effect of electric field on ferroelectric properties of BDFO/ZnO thin films, electric polarization measurements were carried out using ferroelectric hysteresis loop tracer Precision LC from Radiant Technology, USA at a frequency 100 Hz. Platinum was used as the bottom electrode while gold pads served as the top electrode for the polarization measurements. Ferroelectric hysteresis loop (P-E) obtained at room temperature for thin films grown on Si are shown in Fig. 9. Initially, minimal required voltage was applied to start the measurements. As the electric field increases, there is an improvement in ferroelectric hysteresis loops. A well saturated ferroelectric hysteresis loop having polarization (Ps) of ~8.5 μC/cm$^2$, remnant polarization (Pr) of ~2.6 μC/cm$^2$ and coercive field (E$_C$) of 28 kV/cm for maximum applied electric field of 50 kV/cm was observed. In BDO it is generally accepted that Ec is about 110 kV/cm. However, ferroelectric measurements in this study suggested a much smaller Ec and Pr.
Basically, the electric field required for ferroelectric domain alignment is known to depend on intrinsic properties of the material, presence of impurities in the sample, grain size and orientation, defects in grain boundaries, and stresses [12]. The co-existence of magnetic and ferroelectric properties has been observed in ZnO-BDFO films on Si, in order to have the flexibility in device design.

### 4. CONCLUSION

The BDFO/ZnO films are directly deposited on Si substrate exhibit coexistence of ferromagnetic and ferroelectric ordering with significant coupling at room temperature.

There is also saturation in ferroelectric and magnetic hysteresis loop at room temperature, and magnetization achieved at low value of magnetic field. The trend in variation of magnetic as well as low temperature properties could be ascribed to stress induced during the growth process. The coexistence of ferroelectric and magnetic domains in the spatial area of a few microns in these thin films has been established by using multimode atomic force microscopy (MAFM). The results suggests their excellent multifunctional properties, BDFO/ZnO films/composite could be significantly useful during realization of devices in novel applications such as MEMS devices, memory, sensors, electricity generators, and strong history dependent devices etc.

### ACKNOWLEDGEMENT

The authors wish to acknowledge partial funding received from the Department of Information Technology, Government of India, through the Centre of Excellence in Nanoelectronics, IIT Bombay (Grant No. 08DIT006). The authors also wish to thank Department of Physics and Department of Material Science, IIT Bombay for experimental help.

*Email:* $^{b)}$himsharma@iitb.ac.in
$^{a)}$ keshav_eck26@yahoo.com

Email: [b)]himsharma@iitb.ac.in
[a)] keshav_eck26@yahoo.com

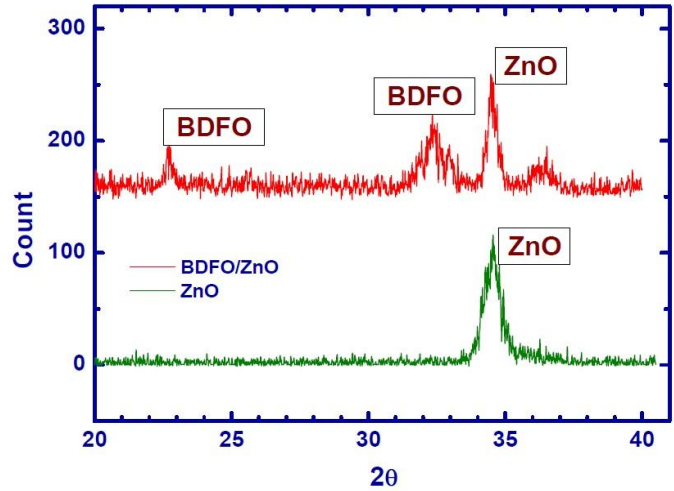

**Fig.1.** X-ray diffraction pattern obtained of BDFO/ZnO films grown on Si substrate

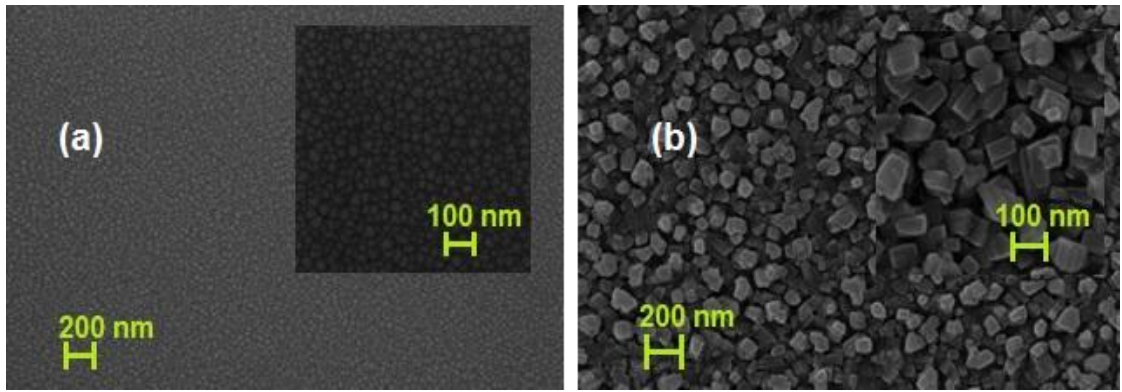

**Fig.2.** (a) SEM image of ZnO Layer at 200nm. Inset shows the image at 100nm (b) SEM image of BDFO/ZnO Layer at 200nm. Inset shows the image at 100nm

*Email:* [b)]*himsharma@iitb.ac.in*
[a)] *keshav_eck26@yahoo.com*

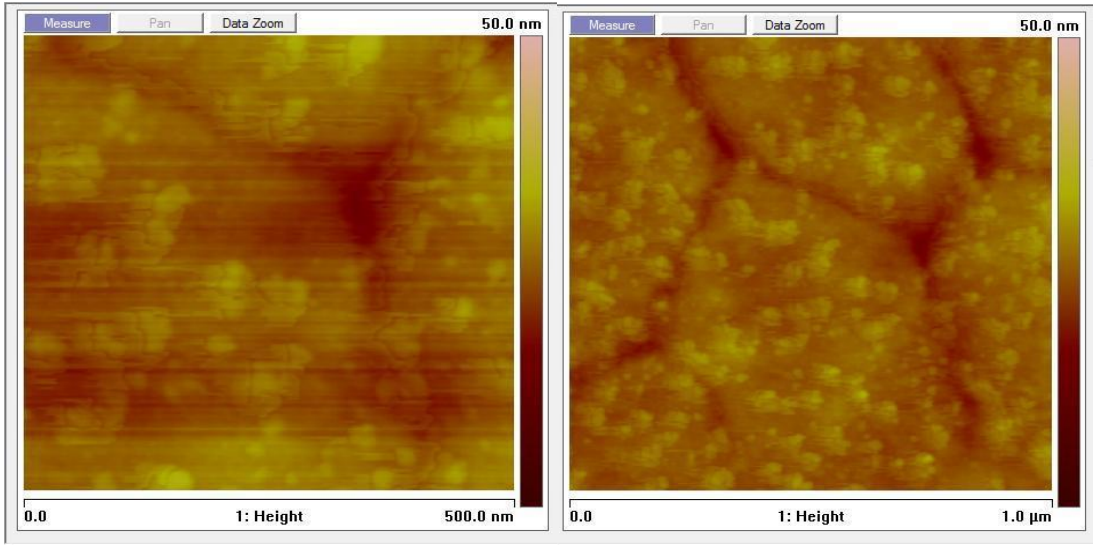

**Fig.3.** AFM (Topography) Images of BDFO thin films of 300nm for scan height (a) 500 nm (b) 1μm

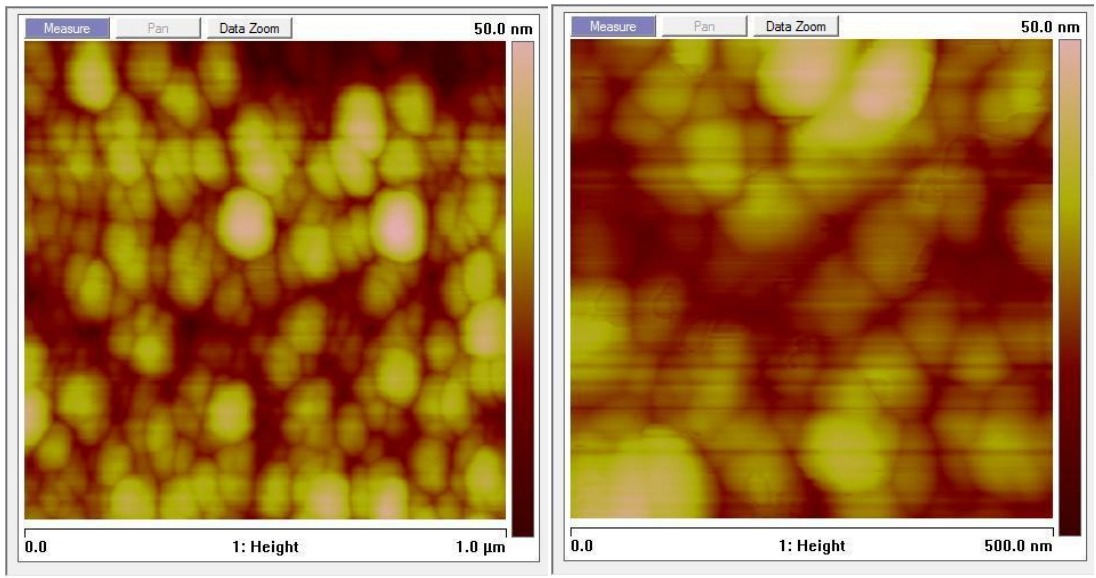

**Fig.4.** AFM (Topography) Images of BDFO/ZnO thin film of 300 nm for scan height (a) 500 nm (b) 1μm.

*Email:* [b)himsharma@iitb.ac.in](mailto:himsharma@iitb.ac.in)
[a)keshav_eck26@yahoo.com](mailto:keshav_eck26@yahoo.com)

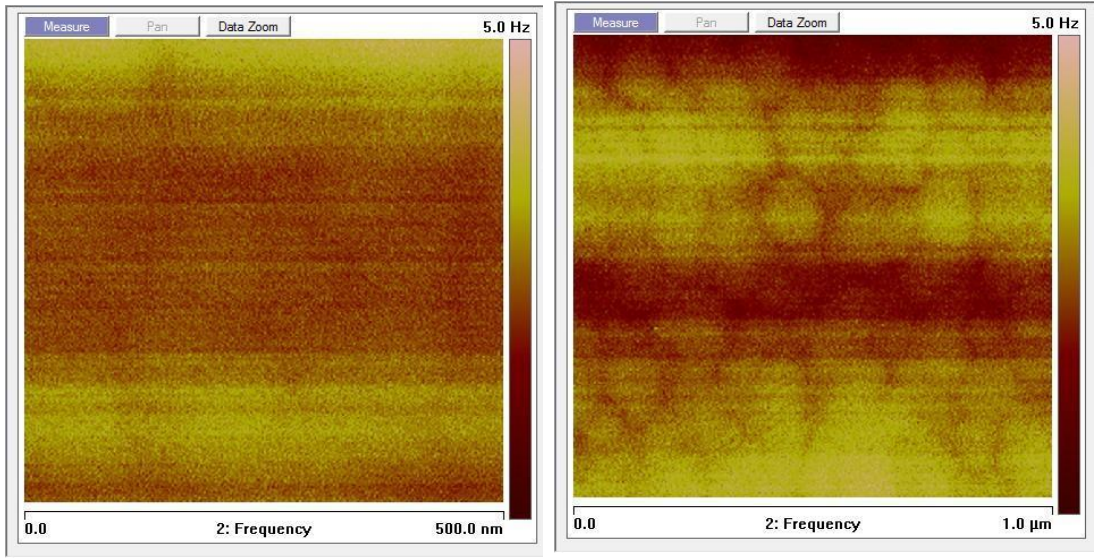

**Fig.5.** MFM Images of BDFO/ZnO thin films of 300nm for scan area (a) 500nm (b) 1μm

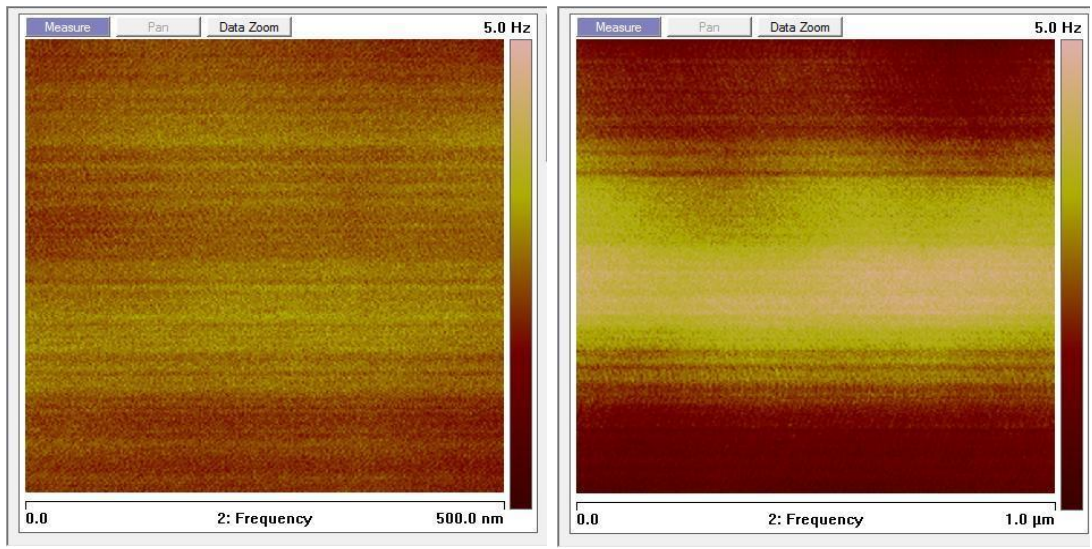

**Fig.6.** MFM images of BDFO thin films of 300nm thickness for scan area (a) 500nm (b) 1μm

Email: [b)]himsharma@iitb.ac.in
[a)]keshav_eck26@yahoo.com

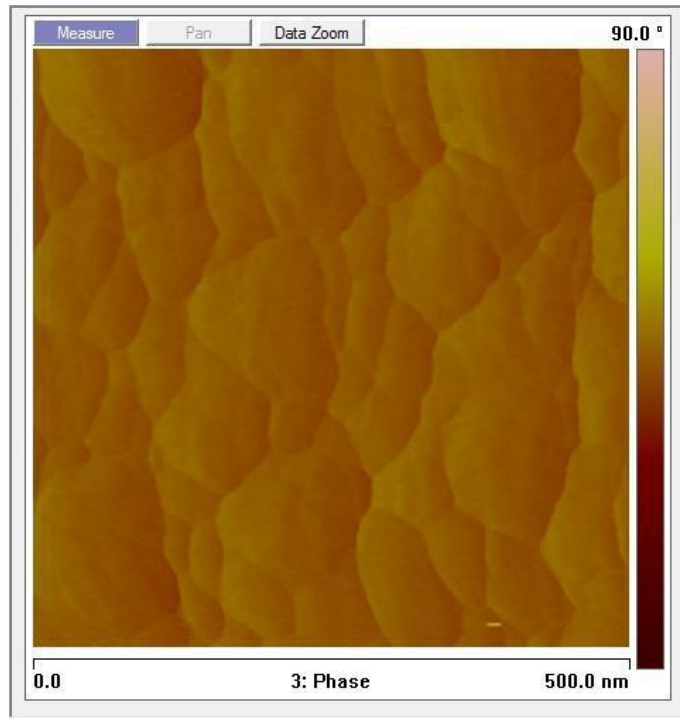

**Fig.7.** Phase Image of BDFO/ZnO thin films of 300nm thickness for scan area 500nm

*Email:* [b)himsharma@iitb.ac.in](mailto:himsharma@iitb.ac.in)
[a)keshav_eck26@yahoo.com](mailto:keshav_eck26@yahoo.com)

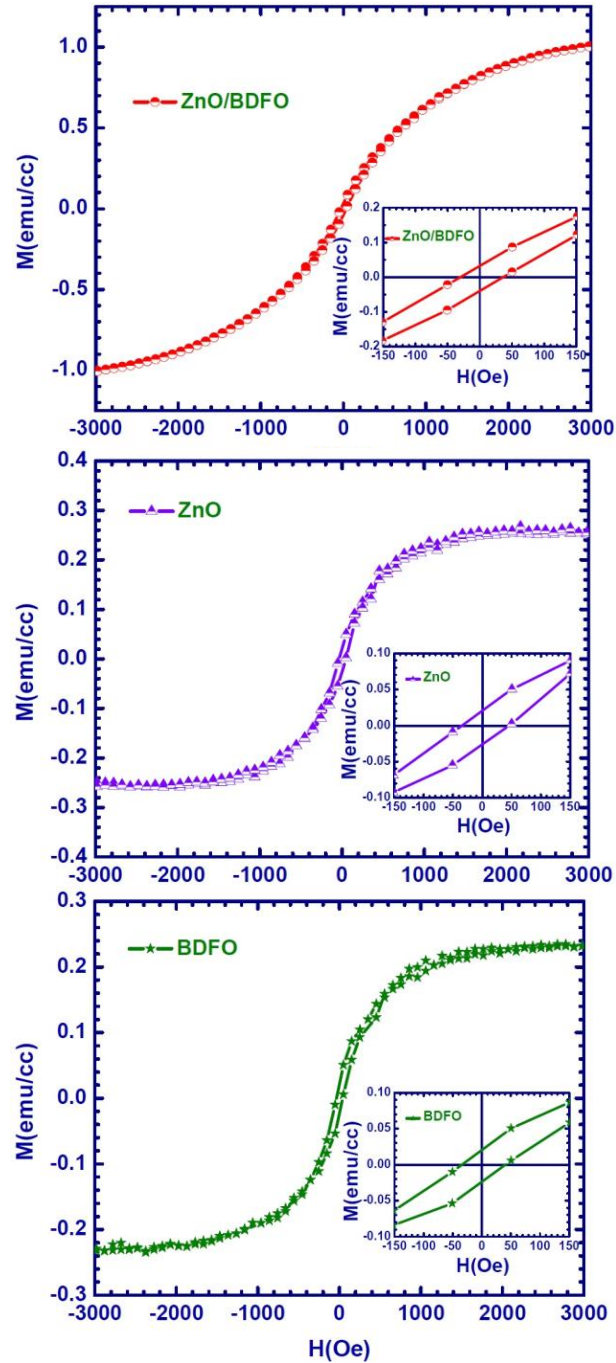

**Fig.8.** Magnetization with variation in applied magnetic field (M-H) curves obtained for $Bi_{1-x}Dy_xFeO_3$(BDFO) and ZnO thin films grown on Si substrate using S-VSM (a) ) BDFO/ZnO (b) ZnO (c) BDFO

*Email:* [b)]himsharma@iitb.ac.in
[a)]keshav_eck26@yahoo.com

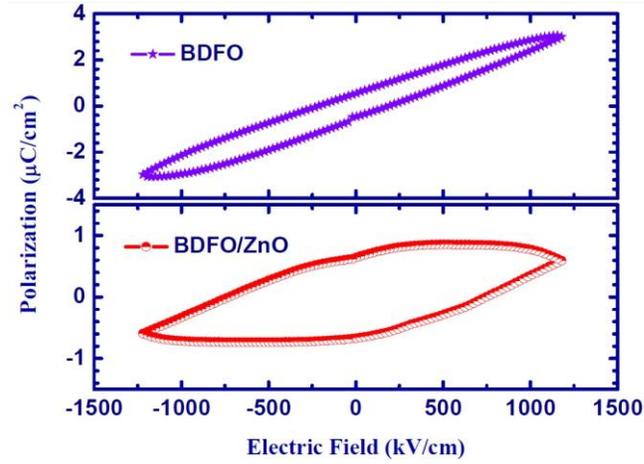

**Fig.9.** Polarization Vs Electric field (P-E) curves obtained for $Bi_{1-x}Dy_xFeO_3$ (BDFO) and BDFO/ZnO thin films grown on Si substrate at room temperature.

*Email:* [b)]*himsharma@iitb.ac.in*
[a)]*keshav_eck26@yahoo.com*